\documentclass[useAMS,usenatbib]{mn2e}

\usepackage{graphicx}
\usepackage{array,longtable}
\usepackage{journals}
\usepackage{url}

\usepackage{latexsym}

\usepackage{ulem}

\usepackage{color}

\title[Large scale nested stellar discs in NGC~7217]
{Large scale nested stellar discs in NGC~7217\thanks{Based on the observations with the Russian 6m telescope}}
\author[Sil'chenko et al.]{Olga K. 
Sil'chenko$^{1,2}$\footnotemark[1], 
Igor V. Chilingarian$^{3,1}$, Natalia Ya. Sotnikova$^{4,5}$\thanks{E-mail: 
nsot-astro@mail.ru (NYS); olga@sai.msu.su (OKS)}, 
\and Victor L. Afanasiev$^{6}$\\
$^{1}$Sternberg Astronomical Institute, Moscow State University, 13
Universitetski prospect, 119992, Moscow, Russia\\
$^{2}$Isaac Newton Institute of Chile, Moscow Branch\\
$^{3}$Centre de Donn\'ees astronomiques de
Strasbourg -- Observatoire de Strasbourg, CNRS UMR~7550, \\ Universit\'e de
Strasbourg, 11 Rue de l'Universit\'e, 67000 Strasbourg, France\\
$^{4}$St.Petersburg State University, Russia\\
$^{5}$Isaac Newton Institute of Chile, St. Petersburg Branch\\
$^{6}$Special Astrophysical Observatory, Russian Academy of Sciences, Nizhnij Arkhyz, Russia}
\begin{document}

\date{Accepted 2011 March 7. Received 2011 March 2; in original form 2010
November 4}

\pagerange{\pageref{firstpage}--\pageref{lastpage}} \pubyear{2011}

\maketitle

\label{firstpage}

\begin{abstract} 
NGC~7217 is an unbarred early-type spiral galaxy having a multi-segment
exponential light profile and a system of starforming rings of
the unknown origin; it also possesses a circumnuclear gaseous polar disc. 
We analysed new long slit spectroscopic data for NGC~7217 and derived
the radial distributions of its stellar population parameters and stellar and gaseous
kinematics up to the radius of $r\approx 100\arcsec$ ($\sim 8$ kpc). We
performed the dynamical analysis of the galaxy by recovering its velocity
ellipsoid at different radii, and estimated the scaleheights of its two
exponential discs. The inner exponential stellar disc of NGC~7217
appears to be thin and harbours intermediate age stars
($t_{\rm{SSP}}\approx5$~Gyr). The outer stellar disc seen between the radii
of 4 and 7~kpc is very thick ($z_0 = 1\dots3$~kpc), metal-poor,
[Fe/H]$<-0.4$ dex, and has predominantly young stars, $t_{SSP}=2$~Gyr. The
remnants of minor mergers of gas-rich satellites with an early-type giant
disc galaxy available in the GalMer database well resemble different
structural components of NGC~7217, suggesting two minor merger events in the
past responsible for the formation of the inner polar gaseous disc and large
outer starforming ring. Another possibility to form the outer ring is the re-accretion
of the tidal streams created by the first minor merger.
\end{abstract}


\section{Introduction}

Currently, our views on galaxy structure are changing dramatically. The
classical view that every disc galaxy represents a combination of a
de-Vaucouleurs' bulge, a sort of an elliptical galaxy inside a larger
stellar system, and an exponential large scale stellar disc (e.g.
\citealp{Freeman70}) does not conform to modern high accuracy observations,
in particular to photometric data. With the surface photometry reaching low
surface brightness limits, down to 27--28~mag~arcsec$^{-2}$ in the $r$-band,
the brightness profiles of most large scale stellar discs cannot be fitted
by a single component exponential law: 90~per~cent of discs turn to be
either truncated or antitruncated \citep{PT06}. Bulges also appear to be far 
from de-Vaucouleurs' spheroids: when approximated by a Sersic law with a
free power parameter, they show surface brightness profile shapes spanning
a range of $n$, with the mean $n\approx 2$ 
\citep{APB95,SJ98,Graham01,MH01}. 
Some morphological types, namely,
spirals later than Sbc \citep{AS94} and lenticulars \citep{BGDP03,LSB05},
exhibit exponential surface brightness profiles for their bulges. The
scatter in the concentration parameter $n$ is thought to be related to variety 
of bulge formation mechanisms; a range of scenarios restricts also dynamical 
properties of the bulges of different types. 

Currently, all bulges are divided into two categories, ``classical''
high luminosity bulges which are suggested to be formed by fast violent
events like mergers, and ``pseudobulges''. The bulges of the latter type
promulgated by J.~Kormendy (see \citealp{Kormendy93} and earlier conference
contributions), are thought to form by secular evolution from the gaseous
and stellar material of the disc \citep{KK04}, in particular through the
central disc heating by bars \citep{CS81}, hence they have to resemble
discs by some of their dynamical characteristics. 
\citet{BGP07} noted a presence of a lot of separate discy stellar 
structures in the centers of spiral galaxies, and 
\citet{EBGB03}
demonstrated that sometimes inner discs in early-type disc galaxies might
resemble bulges. Although, following this fashion, a great variety of
central structures often found in early-type disc galaxies \citep[see
e.g.][]{ES02} is now treated as the pseudobulges including nuclear discs
and bars \citep{DF07}, we would like to stay on the classical point of view
that the principal difference between discs and bulges is related to their
thickness: discs are thin, flat, roughly two-dimensional structures, and
bulges must be thick and three-dimensional by definition. Pseudobulges being
produced by secular-evolution mechanisms may have exponential surface
brightness profiles \citep{FB95}, but they differ principally from the
exponential stellar discs by their thickness. To put a quantitative
criterion to distinguish between discs and spheroids, let us take a look on
the phenomenology. E.~Hubble classified elliptical galaxies by their shape
ranged between \textit{E0} (the axis ratio of 1) and \textit{E7} (the axis
ratio of 3); however many former \textit{E7} galaxies are now thought to be
lenticulars, NGC~3115 being a famous example. As for the large-scale stellar
discs, the closest axis ratio to the transition towards spheroids is perhaps
demonstrated by the thick disc of the edge-on spiral galaxy NGC~891: its
ratio of the exponential scalelength to the exponential scaleheight is about
3.3 \citep{IMR09}. Hence, we consider that the ratio of scalelength to
scaleheight of about 3 is a reasonable frontier between spheroids and
discs.

While analysing a photometric structure of a disc galaxy inclined to the 
line of sight, we can trace the ellipticity of its isophotes to establish a 
transition radius where the thick central structure, the bulge, delegates a 
dominance to the thin disc: it is the radius where the isophote ellipticity 
stops raising.
For a galaxy at a small inclination, it is very difficult to 
distinguish between its exponential pseudobulge and a disc without a 
3D dynamical model of a galaxy. We attempted to construct such a model 
for the galaxy NGC 7217, for which there is still no agreement on the 
structure and origin of its inner subsystems.

NGC~7217 is a giant early-type spiral galaxy seen almost face-on. The
distance to the galaxy adopted in our paper, 18.4~Mpc, was estimated 
from the Tully--Fisher relation by
\citet{Russell02}. It corresponds to the spatial scale of
0.08~kpc~arcsec$^{-1}$. NGC~7217 is listed in the catalogue of isolated
galaxies by \citet{Karachentseva73}; in a more recent study by
\citet{BGP01}, the density of its environment is also estimated as zero. The
galaxy whose evolution is supposedly free of the environmental influence,
demonstrates a set of enigmatic structures which can be best explained by (a
set of) minor mergers. First of all, it has three starforming rings, at
radii of 11, 33, and 75~arcsec \citep{BC93} looking like resonance
structures requiring the presence of a non-axisymmetric potential
\citep{VBA95}, while the galaxy itself is unbarred. \citet{Buta+95} proposed
to decompose the whole galaxy into a small disc and a large mildly triaxial
de-Vaucouleurs spheroid dominating at all distances from the center. The
dominance of the bulge triaxial potential over the whole galaxy caused the
formation of three resonance rings with high density gas concentration and
ongoing star formation. Later we undertook our own decomposition of the
NGC~7217 structure and have found the two large exponential components with
different 
scalelengths\footnote{The bulge also demonstrates an exponential profile 
with a scalelength $\sim$ 4 arcsec \citep{SA00}.}, 
12.5 arcsec, or 1~kpc for the inner one and 
35.8 arcsec, or about 3 kpc for the outer one 
\citep{SA00}. The inner component has larger intrinsic
ellipticity than the outer one and so may be oval providing the necessary
triaxiality of the potential to put the rings at the resonance radii. The
bulge, if any, is small, confined within the innermost (nuclear) ring, and
exponential. To choose between the alternate ways of decomposition, the
kinematical data and a dynamical model are needed. The situation is
complicated by the fresh (?) remnants of a minor merger presented by
counterrotating stars in the inner disc \citep{MK94} and the inner gas polar
disc within $R\approx 4$~arcsec \citep{SA00}.

In this paper we present the results of long slit spectroscopy of NGC~7217.
Our goal is the diagnostics of the nature of the two large scale exponential
structures in this galaxy. The inner exponential structure may be an inner
disc (a thin structure) or a pseudobulge (a thick structure). Since NGC~7217
is seen nearly face-on, the isophote ellipticity is a poor indicator of the
stellar component thickness as it is close to zero over the whole galaxy.
But with the stellar velocity dispersion profiles along the major and minor
axes we can try to estimate the scaleheights of the stellar components from
dynamical considerations. In Section~2 we describe our observations,
data reduction and analysis, in Section~3 we provide our new estimates of
the stellar population properties at different radii. The principal
Section~4 gives our dynamical consideration of the NGC~7217 large scale
structures, and Section~5 contains a discussion about the origin
of the galaxy structure including the comparison with numerical simulations.
In Section~6 we give a brief summary of our results.

\section{Observations and Data Reduction.}

The spectroscopic observations were carried out with the
SCORPIO\footnote{For a description of the SCORPIO instrument, see
\url{http://www.sao.ru/hq/moisav/scorpio/scorpio.html}} universal
spectrograph \citep{AM05} installed at the prime focus of the Russian 6-m
Bol'shoy Teleskop Azimutal'nyy (BTA) operated by the Special Astrophysical
Observatory, Russian Academy of Sciences. We used the VPHG2300G grating
providing an intermediate spectral resolution ($R \approx 2200$) in a
relatively narrow wavelength region ($4800 < \lambda < 5500$~\AA) however
containing a rich set of strong absorption line features making it suitable
for studying both internal kinematics and stellar populations of a galaxy.
The chosen spectral range also includes several emission lines, H$\beta$,
[O{\sc{iii}}], and [N{\sc i}], which we used to derive the gas kinematics
and line ratios. The
slit was 1.0~arcsec wide and 6~arcmin long. The 2k$\times$2k EEV CCD42-40
detector used in the 1$\times$2 binning mode provided a spectral sampling of
0.37~\AA~pix$^{-1}$ and a spatial scale of 0.357~arcsec~pix$^{-1}$.

We observed NGC~7217 in two slit positions going through the centre,
along minor and major axes of its inner isophotes. The major axis spectrum,
P.A.=81~deg, was obtained on 6/Oct/2008 with an integration time of 80~min
under atmosphere conditions with good transparency and intermediate image
quality of $1.7\arcsec$ FWHM corresponding to the median seeing at the
telescope site. The minor axis data at P.A.=169~deg were collected during
two nights October, 6, 2008 and October, 8, 2008 with a total exposure time of 75~min
under bad variable transparency and seeing quality of $\sim3$~arcsec making
them notably shallower than the major axis spectra. We obtained the
following calibrations: night time internal flat field and arc line spectra,
the GD~248 spectrophotometric standard star, and high signal-to-noise
twilight spectral frames.

Data reduction and analysis for the spectral data of NGC~7217 was
identical to that of the lenticular galaxy NGC~6340 presented in
\citet{Chilingarian+09b}. We refer to that paper for all details, here we
give only essential information required for understanding our data analysis
and interpretation.

The primary data reduction steps included bias
subtraction, flat fielding and cosmic ray hit removal using the Laplacian
filtering technique \citep{vanDokkum01}.  Then, we built the wavelength
solution by identifying arc lines and fitting their positions using a 3rd
order two-dimensional polynomial along and across dispersion, and linearised
the spectra. The obtained wavelength solution had fitting residuals of about
0.08~\AA~RMS.

The SCORPIO spectrograph has significant variations of the spectral
line-spread-function (LSF) along and across the wavelength direction (see
\citealp{Moiseev08} and \citealp{Chilingarian+09b} for details). In our
observations of NGC~7217, we used the peripheral regions of the slit beyond
2~arcmin from the galaxy centre to estimate the night sky spectrum,
therefore it was very important to take the LSF variations into account in
order to minimize the effects of the sky subtraction artefacts on the data
analysis. We mapped the LSF by fitting the high-resolution ($R=10000$) Solar
spectrum against the twilight spectra at 64 positions along the slit in five
slightly overlapping wavelength segments covering the spectral range of the
SCORPIO setup with the penalized pixel fitting technique \citep{CE04}. We
used the Gauss-Hermite parametrization up to the 4th order \citep{vdMF93} to
represent the LSF shape.

Then, we modelled the night sky spectrum at every position along the
slit following the procedure described in detail in \citet{Chilingarian+09b}
where the main step was the parametric signal recovery applied to the sky spectrum
taken in the outer regions of the slit in the Fourier space as follows:
\begin{equation}
f(x, \lambda) = F^{-1}(F(f(\mbox{sky}, \lambda))
\frac{F(\mathcal{L}(x))}{F(\mathcal{L}(\mbox{sky}))}),
\end{equation}
\noindent where $f(x, \lambda)$ denotes a sky spectrum at the position $x$
along the slit with its parametrized LSF $\mathcal{L}(x)$; $f(\mbox{sky},
\lambda)$ is the night sky spectrum in any region of the slit with the LSF
$\mathcal{L}(\mbox{sky})$, and $F$, $F^{-1}$ are the direct and inverse
Fourier transforms respectively. The night sky model created in this fashion
results in a nearly Poisson quality of the sky subtraction which is
absolutely crucial for our analysis of the external regions of the galaxy.

\section{Stellar Populations and Internal Kinematics of NGC~7217}

We derived the parameters of internal kinematics and stellar
populations of NGC~7217 by fitting high-resolution {\sc pegase.hr}
\citep{LeBorgne+04} simple stellar population (SSP) models against our
spectra with the {\sc NBursts} full spectral fitting technique
\citep{CPSA07,CPSK07}. The SSP models were computed using the
\citet{Salpeter55} stellar initial mass function. The fitting algorithm is
picking up a template from a grid of stellar population models in the
age--metallicity space convolved with the instrumental response of the
spectrograph (see previous section), broadening it with the line-of-sight
velocity distribution (LOSVD) of a galaxy represented by the Gauss-Hermite
parametrization up to the 4th order, i.e. $v$, $\sigma$, $h_3$, and $h_4$.
The models are multiplied pixel by pixel by the $n^{\rm{th}}$ order Legendre
polynomial continuum which is used to account for possible imperfections of
the flux calibration and for the internal dust extinction in a galaxy.
All kinematical and stellar population parameters are determined in
a single non-linear minimization loop therefore reducing possible
degeneracies. The $\chi^2$ is penalized towards purely Gaussian LOSVD as
explained in \citet{CE04} in order to stabilise the solution in case of low
signal-to-noise ratios and/or insufficient spectral sampling of the data.
We stress that for the dynamical analysis presented below, we use the pure
Gaussian LOSVD parametrization without $\chi^2$ penalization.

The {\sc pegase.hr} models are constructed from empirical stellar
spectra in the Solar neighbourhood where the value of $\alpha$-element
abundances is known to correlate with the overall metallicity. Therefore, at
intermediate and high metallicities, our SSP models are representative of
the solar $\alpha$/Fe abundance ratio. However, \citet{Chilingarian+08}
showed that supersolar $\alpha$/Fe ratios bias neither age nor metallicity
estimates when using the {\sc NBursts} spectral fitting technique, and
\citet{SA00} found nearly solar $\alpha$/Fe in NGC~7217.

NGC~7217 possesses quite strong emission lines in its spectrum
especially in the regions corresponding to the rings with ongoing star
formation. Therefore we had to exclude from the fitting procedure several
narrow 20~\AA--wide regions around emission lines (H$\beta$ $\lambda =
4861$~\AA, [O{\sc iii}] $\lambda = 4959, 5007$~\AA, and [N{\sc i}] $\lambda
= 5197, 5201$~\AA) redshifted according to the line-of-sight velocity of NGC~7217.
\citet{Chilingarian09} demonstrated that H$\beta$ contains 20~per~cent of
the age-sensitive information at maximum when using the {\sc NBursts}
technique in a spectral range similar to ours, therefore excluding it from
the fit neither biases age estimates (see also Appendix~A2 in
\citealp{CPSA07} and Appendix~B in \citealp{Chilingarian+08}), nor degrades
significantly the quality of the age determination.

We fitted Gaussians pre-convolved with the SCORPIO LSF into the [O{\sc
iii}] and H$\beta$ emission lines in the residuals of the stellar
population fitting and determined the ionised gas kinematics independently in
the two lines as well as the emission line flux ratios.

\begin{figure}
\centering
\includegraphics[width=0.65\textwidth,angle=270]{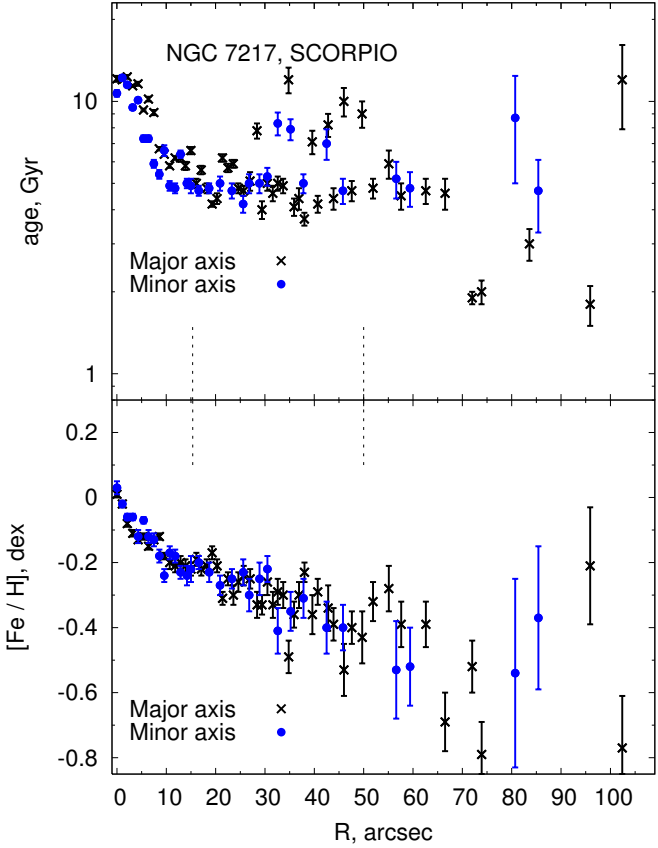}
\caption{Radial distributions of the mean age (top) and metallicity (bottom) 
of the stellar populations measured along the major ($PA=81^{\circ}$) and 
minor ($PA=169^{\circ}$) axes. The vertical dashed lines indicate the 
conventional boundaries of the bulge and the inner exponential 
component set at four scalelenghts of each component.
\label{stelpop}}
\end{figure}

The obtained radial distributions of SSP-equivalent age and metalliticity of
NGC~7217 are shown in Fig.~\ref{stelpop}. \citet{SA00} analysed the large
scale structure of NGC~7217 and identified three galaxy components, a bulge
dominating at radii $5 - 15$~arcsec (0.4--1.2~kpc), and two exponential
structures interpreted as discs. The inner and outer discs dominate in the
galaxy light profile at $20 - 50$~arcsec (1.6--4~kpc) and $60 - 110$~arcsec 
(4.8--8.8~kpc) respectively.

The age profile of NGC~7217 presented in Fig.~\ref{stelpop} exhibits
specific features at the radii dominated by these three substructures. In
the bulge dominated region located mostly inside the nuclear starforming
ring ($R=10 - 12$~arcsec), the mean stellar age decreases from 10--13~Gyr in
the centre to 5~Gyr outwards. In the inner 
exponential component 
the age stays nearly constant at about 5~Gyr; individual estimates at the radii 
between 20 and 50~arcsec have rms of 0.7~Gyr in the major-axis profile and 
0.9~Gyr in the minor-axis one. The outer 
exponential component 
($R=60 - 110$~arcsec) has an intermediate age of about 2--3~Gyr, whereas in 
the broad outer starforming ring ($R=70 - 80$~arcsec) it drops down to 1~Gyr. 
Surprisingly, at these radii we also see a drop in the SSP-equivalent stellar 
metallicity down to $-0.65$~dex prominent in the eastern side of the galaxy 
but hardly detected in the western side, which is probably connected to the 
morphology and location of individual starforming regions crossed by the slit. 
The mean metallicity of the bulge is close to the solar value, 
[Fe/H]=$-$0.06~dex, while the outer 
exponential component 
([Fe/H]=$-$0.45~dex) is rather metal poor for such a massive galaxy. The inner 
exponential component 
has a mean metallicity of about $-0.2$~dex with a notable metallicity gradient 
around $-0.3$~dex per dex in radius.

Hence, the existence of the three components found in the
photometric data by \citet{SA00}, a bulge and two [probably] exponential
discs, is supported by specific features in the stellar population profiles.
Later, we will propose the evolutionary scenario of NGC~7217, where the two
discs may have different origin.

\begin{figure}
\centering
\includegraphics[width=0.35\textwidth,angle=270]{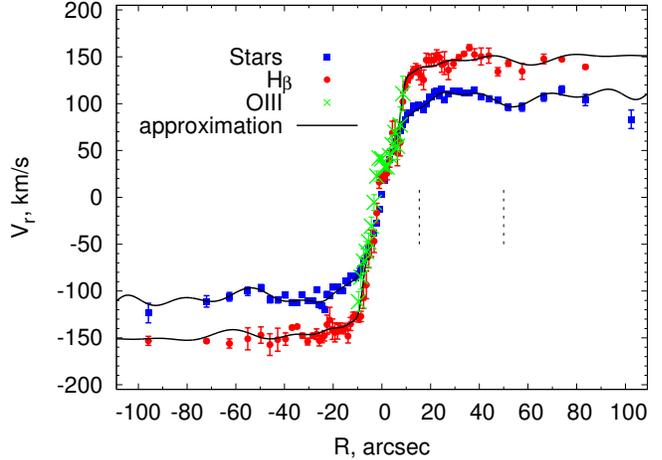}
\caption{Radial distribution of the line-of-sight velocities of stars and
ionized gas along the major ($PA=81^{\circ}$) axis. 
The [O{\sc iii}] data are presented only for $|R| < 10$~arcsec. 
The vertical dashed lines indicate the conventional boundaries of the 
bulge and the inner exponential component (see Fig.~\ref{stelpop}).}
\label{losvel}
\end{figure}

\begin{figure}
\centering
\includegraphics[width=0.65\textwidth,angle=270]{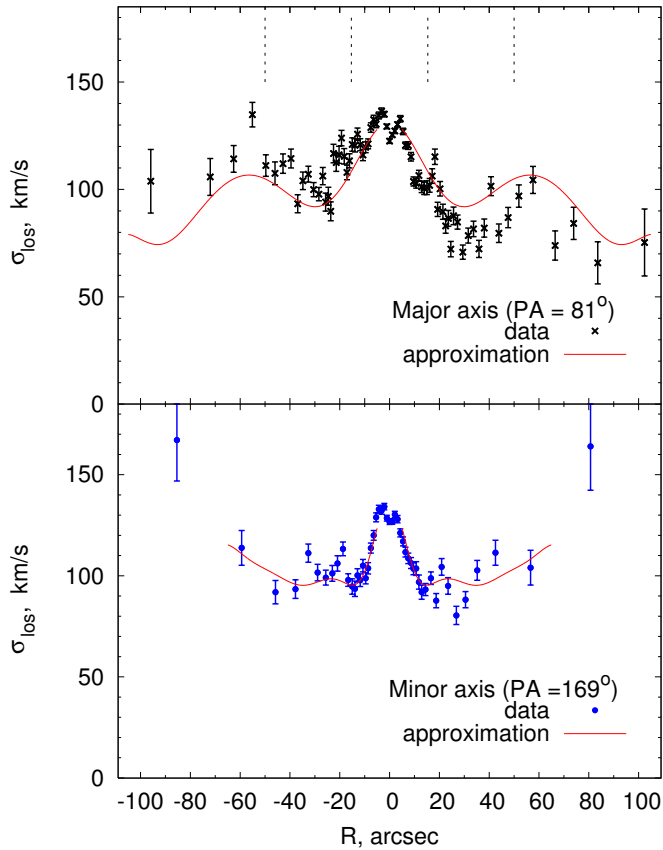}
\caption{Radial distributions of the line-of-sight stellar velocity dispersion
measured along the major, $PA=81^{\circ}$, (top) and minor, 
$PA=169^{\circ}$, (bottom) axes. 
The vertical dashed lines indicate the conventional boundaries of the 
bulge and the inner exponential component (see Fig.~\ref{stelpop}).}
\label{lossigma}
\end{figure}

We present the derived major axis (P.A.=81~deg) profiles of
line-of-sight velocities for the gaseous and stellar components in
Fig.~\ref{losvel}. The H$\beta$ emission line component remains clearly visible
along the whole extent of the galaxy where we can measure the stellar
kinematics. The [O{\sc iii}] is much weaker than H$\beta$ at radii $R >
10$~arcsec. The velocities obtained from the two emission lines agree
well, but those determined from H$\beta$ in the outer region have 4--5 times
smaller uncertainties. Therefore, we constructed the combined gaseous line-of-sight
velocity profile displayed in Fig.~\ref{losvel} from [O{\sc iii}] and
H$\beta$ kinematics at radii below and above 10~arcsec correspondingly.
Both, stellar and gaseous velocity profiles have similar shapes including
even some particular features such as a notable drop at $R\approx
50$~arcsec. They are also very symmetric. Therefore, for the dynamical
analysis presented in the next section, we folded them along the galaxy
centre and averaged the values of rotation velocities from both sides. The
gaseous rotation velocities ($\sim 150$~km~s$^{-1}$) exceed the stellar ones 
($\sim 120$~km~s$^{-1}$) by nearly 30~km~s$^{-1}$ at all radii, 
which is an observational manifestation of the asymmetric drift. The difference stays 
nearly constant between 10 and 70~arcsec suggesting similar dynamical statuses 
of the inner and outer exponential stellar structures. These structures are 
probably discs, but quite high asymmetric drift implies that they should be 
rather thick.

In Fig.~\ref{lossigma} we show the observed radial distributions of the
line-of-sight stellar velocity dispersion along minor and major axes of
NGC~7217. Since the galaxy is seen nearly face-on, the vertical ($\sigma_z$)
component of the disc velocity ellipsoid creates the main contribution
to the observed velocity dispersion. However, there is a notable difference
between major- and minor-axis velocity dispersion profiles suggesting
significant contributions of the tangential ($\sigma_{\varphi}$) and radial
($\sigma_R$) velocity dispersions to the values along the major and minor
axes respectively. The main feature of the stellar line-of-sight velocity
dispersion profiles of NGC~7217 is a minimum in the area of the inner
exponential component and a smooth raise outwards. This behaviour is quite
unexpected: as we demonstrated, the outer stellar component is younger than
the inner one, so it would be more natural if the stellar subsystem formed
recently from the dynamically cold gas is itself dynamically cold. However,
the velocity dispersion rise beyond $R \approx 50$~arcsec is
statistically significant and seen in both minor- and major-axis velocity
dispersion profiles. In Section~5 we will give the possible explanation 
of these features.

\section{Dynamical Analysis: Two Very Different Discs in NGC~7217}

We used our kinematical data to recover the stellar velocity
ellipsoid and reconstruct the radial distributions of all stellar velocity
dispersion components. Then, we calculated  the stellar disc thickness
profile from the $\sigma_z$ radial distribution. By using relations 
describing the disc equilibrium,  we did it independently from 
the major- and minor-axis kinematical data.

For an intermediately inclined galaxy, the two in-plane components of the
stellar velocity dispersion, $\sigma_R$ and $\sigma_{\varphi}$, are related
to the observed minor- and major-axis line-of-sight velocity dispersions
according to the following relations:
\begin{equation}
\begin{array}{rcl}
\sigma_\mathrm{los,min}^2 (R \cos i) & = 
& \sigma_R^2 \, \sin^2 i + \sigma_z^2 \, cos^2 i \, ,\\
& &\\
\sigma_\mathrm{los,maj}^2 (R) & = 
& \sigma_\varphi^2 \, \sin^2 i + \sigma_z^2 \, cos^2 i \, .
\end{array}
\label{slos}
\end{equation}
They also include the contribution from the vertical component of 
the velocity ellipsoid $\sigma_z$.
Thus, by measuring the variations in the line-of-sight dispersions 
along the principal axes, we can obtain only a linear combination of 
$\sigma_R(R)$, $\sigma_\varphi(R)$ and $\sigma_z(R)$. Therefore, we need 
additional information in order to derive all three velocity dispersion 
components from these equations. To close the system of 
equations~(\ref{slos}), we can use some dynamical relations, which are valid 
if the system is in equilibrium. One such relation connects the
velocity dispersion components $\sigma_R(R)$ and $\sigma_\varphi(R)$
via the mean azimuthal velocity of stars \citep{BT87}
\begin{equation}
\frac{\sigma_\varphi^2}{\sigma_R^2} = \frac{1}{2} 
\left( 1 + \frac{\partial \ln \bar{v}_\varphi}{\partial \ln R}\right) 
\, .
\label{vphi}
\end{equation}
If the rotational velocities are unknown, we can use the local circular 
speed of gas $v_\mathrm{c}$ instead of $\bar{v}_\varphi$ in order to relate 
$\sigma_R ^2$ and $\sigma_\varphi^2$
\begin{equation}
\frac{\sigma_\varphi^2}{\sigma_R^2} = \frac{1}{2} 
\left( 1 + \frac{\partial \ln v_\mathrm{c}}{\partial \ln R}\right) 
\, .
\label{vc}
\end{equation}
This relation is true if most orbits in a disc are quasi-circular.

Adding any of these relations, Eq.~\ref{vphi} or Eq.~\ref{vc}, 
to the two presented above, 
enables us to recover radial distributions of all 
three velocity dispersion components.  For the first time this technique 
was applied to the data for NGC~488 by \citet{GKM97}. However, this
procedure creates very noisy output when applied directly to the data,
because it includes the subtraction of the two quantities
$\sigma_\mathrm{los,min}^2$ and $\sigma_\mathrm{los,maj}^2$ with very close
values, as well as the numerical derivation of $\bar{v}_\varphi$ or
$v_\mathrm{c}$. A possible solution is to parametrize the kinematical
profiles and to find the best fitting solution (see e.g.
\citealp{GKM97,GKM00,SGvdM03}) which will, however, depend on the adopted 
parametrization. For this reason, we adopted the less parametric approach 
as in \citet{NMA08}. We approximated all kinematical profiles using 
polynomials and calculated all quantities including their derivatives 
analytically. But even then, the subtraction of the major axis
velocity dispersion profile from the minor axis one
results in unreliable and ambigious solution. To avoid this, we use 
the asymmetric drift equation. The deprojected major-axis gas rotation is a measure 
of the circular speed $v_\mathrm{c}$, while the deprojected major-axis stellar 
velocities allow us to determine the mean rotational motion $\bar{v}_\varphi$
related to the radial velocity dispersion $\sigma_R$. Therefore, we can obtain 
the $\sigma_R$ profile using the major-axis velocity profiles for gas 
and stars and the asymmetric drift equation \citep{BT87}: 
\begin{equation}
v_\mathrm{c}^2 - \bar{v}_\varphi^2 = 
\sigma_R^2 \left( 
\frac{\sigma_\varphi^2}{\sigma_R^2} -1 - 
\frac{\partial \ln \Sigma}{\partial \ln R} - 
\frac{\partial \ln \sigma_R^2}{\partial \ln R}
\right) \, ,
\label{AsDr}
\end{equation}
where $\Sigma$ is the stellar surface density. Provided that the
stellar mass-to-light ratio stays nearly constant ($(M/L)_{I}=1.86$ in the
Solar units from the {\sc pegase.2} models, see \citealp{FR97}), we can use 
the surface brightness
instead of surface density $\Sigma$ in Eq.~\ref{AsDr}. We assume that the
$I$-band photometric data trace old stellar population whose internal
kinematics we are studying. Because of the logarithmic derivative, the exact
choice of the $M/L$ ratio is not critical. The asymmetric drift equation can
be used directly to obtain the radial velocity dispersion profile, provided
that the ratio between $\sigma_\varphi^2/\sigma_R^2$ is determined from
Eq.~\ref{vphi} or Eq.~\ref{vc}. However, Eq.~\ref{AsDr} requires the
knowledge of the $\sigma_R$ radial gradient. We assume that in the outer
regions of the galaxy between 30 and 70~arcsec, $\sigma_R$ may decline
exponentially as $\propto \exp(-R/h_\mathrm{kin})$ 
and use the last term in the form $2R / h_\mathrm{kin}$, 
where $h_\mathrm{kin}$ is a free parameter.

The choice of a radial velocity dispersion profile exponentially 
declining with the radius is physically motivated. There is a conventional 
assumption that the disc thickness and the $z$-component of the velocity 
dispersion $\sigma_z$ are connected via the vertical equilibrium condition 
for an isothermal layer \citep{Spitzer42}
\begin{equation}
\sigma_z^2 (R) = \pi \, G\, \Sigma(R) \, z_0 \, ,
\label{thick}
\end{equation}
where $z_0$ is the half-thickness of a homogeneous layer.
For a mass-to-light ratio constant with radius, constant $z_0$ and an 
exponential brightness profile it yields:
\begin{equation}
\sigma_z^2 (R) \propto \exp(-R/h)\, ,
\label{s2_z}
\end{equation}
where $h$ is a disc scalelength.

The disc heating theory suggests that the range in the velocity 
anisotropy $\sigma_z / \sigma_R$ is about $0.4-0.8$ \citep{JB90}, 
which is consistent with observations. 
\citet{GKM97,GKM00} and \citet{SGvdM03} 
modelled the data for several 
galaxies and estimated the values for the velocity anisotropy 
$\sigma_z / \sigma_R$ between $0.5$ and $0.7$, similar to the solar 
neighbourhood \citep{DB98}.

Since the ratio $\sigma_z / \sigma_R$ is close to constant, 
Eq.~\ref{s2_z} yields:
\begin{equation}
\sigma_R \propto \exp(-R/2\,h) \, .
\end{equation}
It implies that $h_\mathrm{kin} = 2\,h$. 
This approach does work for our Galaxy \citep{LF89} but in the case of 
NGC~7217 we did not fix the value of 
$h_\mathrm{kin}$ and estimated it 
iteratively by fitting the radial velocity dispersion profile with the 
exponential law and substituting the value of $h_\mathrm{kin}$ in the 
asymmetric drift equation. 
We needed this value to estimate only the last term in 
Eq.~\ref{AsDr}. 
The exact value of this term may affect the final results but the effect 
it quite insignificant. We were changing the range of $R$ used to fit 
$\sigma_R$ by exponential law and did not notice any difference between 
final results. Having derived the $\sigma_R$ radial distribution from the 
Eq.~\ref{AsDr}, we can then compute the radial profile of the vertical 
component of the velocity dispersion $\sigma_z$ using one of the 
Eq.~\ref{slos}. We choose to use both of them to control the reliability of 
our results. 

Finally, we can obtain the thickness profile from Eq.~\ref{thick} which 
is valid for an isothermal layer. 
The assumption about constant velocity dispersion along the
$z$-direction for discs is in agreement with the results of $N$-body
simulations (see e.g. Fig.~2 in \citealp{SR06}). But one should keep in mind
that Eq.~\ref{thick} gives the upper limit for the disc thickness. If there
is a massive dark halo, the disc thickness will be lower for the same value
of $\sigma_z$.

In order to deproject the velocity profiles we need to know the
inclination of the rotation plane to the line of sight. NGC~7217 has a
roundish appearance, however quite high observed (i.e. projected) rotation
velocities suggest that is not seen completely face-on. The isophote
ellipticity analysis assuming the infinitely thin stellar disc provided the
disc inclination estimates in the range of $i = 26 - 28$~deg
\citep{NvdH07,VBA95,SanchezPortal+00}; whereas the kinematical analysis
usually results in slightly higher values $i = 30 - 31$~deg
(see \citealp{Buta+95,Noordermeer+05} for the H{\sc i} map analysis and
\citealp{SM06} for the stellar kinematics). Keeping in mind possible
considerable thickness of the stellar disc, we adopt the inclination value
of $i = 30$~deg. Varying $i$ between 26 and 35~deg changes the resulting
disc thickness estimates by no more than 10~per~cent. 

\begin{figure}
\centering
\includegraphics[width=0.33\textwidth,angle=270]{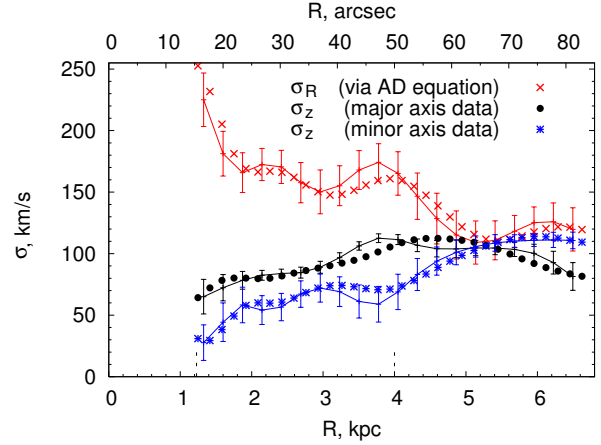}
\caption{Radial distributions of the radial and vertical velocity dispersions
reconstructed from the major- and minor-axis kinematical profiles. 
Full lines -- profiles calculated via the Eq.~\ref{vphi}, 
symbols -- profiles obtained by using the Eq.~\ref{vc}. 
Error bars are shown only for profiles indicated by solid lines.
The vertical dashed lines indicate the conventional boundaries of the 
bulge and the inner exponential component (see Fig.~\ref{stelpop}).}
\label{modsigma}
\end{figure}

We present the reconstructed radial distributions of $\sigma_R$ and $\sigma_z$ 
in Fig.~\ref{modsigma}. 
We estimated uncertainties using a bootstrapping 
method \citep{PTVF92}. We computed $\sigma_R$ and $\sigma_z$
assuming that the errors of observational quantities used in the 
procedure are distributed according to the normal law and changing 
input data points according to these errors. We made several thousands 
realisation of the procedure, averaged resulting values, and estimated 
the dispersion of all the simulations at each point.

\begin{figure}
\includegraphics[width=0.45\textwidth,angle=270]{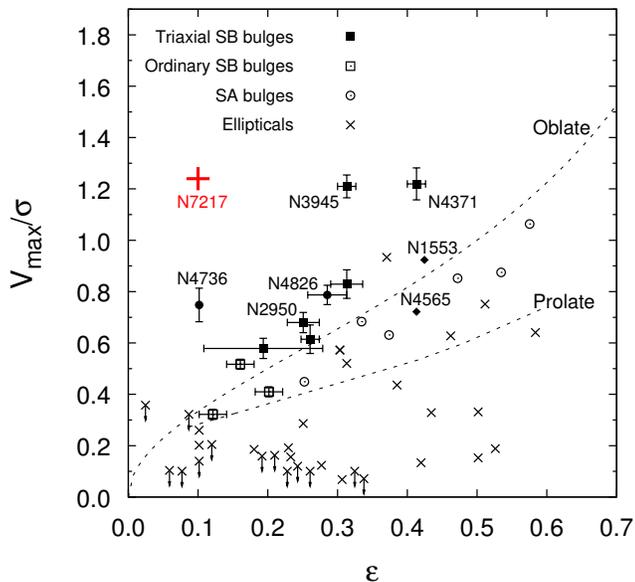}
\caption{Relation between the ratios of observed maximal rotation 
velocities and velocity dispersions and isophote ellipticities for early-type
galaxies, bulges (open symbols) and 
pseudobulges (filled symbols). 
The models of prolate and oblate
isotropic rotators are overplotted by dashed lines. NGC~7217 is shown 
with a red cross being clearly above the locus of pseudobulges from
\citet{KK04}.}
\label{BKfig}
\end{figure}

The data in the Fig.~\ref{modsigma} suggest that the dynamically 
coolest subsystem of NGC~7217 
as concerning the vertical velocity dispersion 
is the {\it inner} stellar exponential component. Therefore it is 
certainly a disc and not a pseudo-bulge. Indeed, in Fig.~\ref{BKfig} we plot the inner 
exponential component of NGC 7217 at the diagram of Binney--Kormendy. With 
its $\epsilon \approx 0.1$ and $v_\mathrm{max}/\sigma _\mathrm{los}=1.24$ 
it stays well above all the pseudobulges dynamically confirmed by Kormendy. 
At $R>30$~arcsec ($>$2.4~kpc) the vertical velocity dispersion component 
slowly raises along outwards reaching 110~km~s$^{-1}$ at the location of the 
outer starforming ring. The $\sigma_R$ profile decreases along the radius as 
expected, but in the transition region where the outer disc starts to dominate 
($R = 40 - 50$~arcsec; 3.2--4.0~kpc), we see a break of the monotonic fall. 
The whole $\sigma_R$ profile can be divided into two parts corresponding to 
the two exponential stellar components.

\begin{figure}
\centering
\includegraphics[width=0.33\textwidth,angle=270]{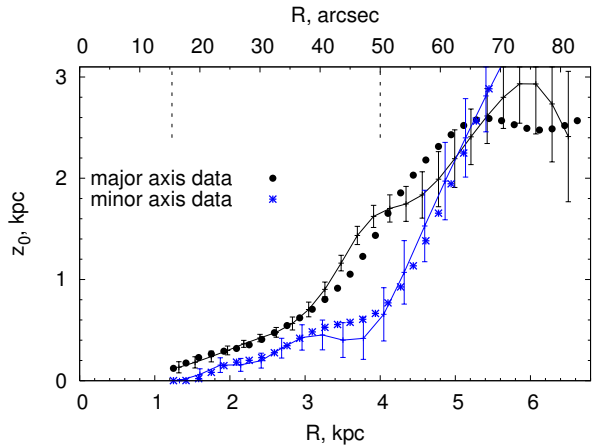}
\caption{Radial distribution of the stellar disc scaleheight -- 
the reconstructed ``edge-on'' view. 
Solid lines are the profiles calculated via Eq.~\ref{vphi}, 
while symbols are for the profiles obtained by using Eq.~\ref{vc}. 
Error bars are shown only for profiles indicated by solid lines.
The vertical dashed lines indicate the conventional boundaries of the 
bulge and the inner exponential component (see Fig.~\ref{stelpop}).}
\label{thickness}
\end{figure}

The disc ``thickness profiles'' at radii $R = 15 \dots 80$~arcsec
reconstructed from the major- and minor-axis kinematics are shown in
Fig.~\ref{thickness}. While the qualitative agreement between them exists,
quantitatively they diverge at some radii. 
As it was mentioned in Section~2, the minor axis data are of a worse
quality than the major axis ones but they are featureless as concerning the 
line-of-sight stellar velocity dispersion. In this sense, both datasets 
are worth each other. One of the sources of the profile
discrepancy might be the adopted inclination angle and the discrepancy 
between measured distances along the major axis and deprojected distances 
along the minor axis.
Similarly to the $\sigma_z$ profile behaviour, we see that the inner stellar 
disc is relatively thin ($z_0=0.2\dots0.7$~kpc). 
Hence, at its inner boundary the ratio of its scalelength to the 
scaleheight is about 4--5. 
The outer disc flares vigorously reaching the half-thickness of 2.5-3~kpc 
at the outer starforming ring position.

\section{Discussion}

\subsection{Explaining the structure of NGC~7217}

From the analysis of the stellar and gaseous kinematics (line-of-sight
velocities and velocity dispersion radial distributions) over an extended
area of the giant early-type disc galaxy NGC~7217, we reconstructed the
velocity ellipsoid and calculated the stellar disc scaleheights at radii
from 20 to 70~arcsec (1.6 to 6~kpc). Earlier, \citet{SA00} demonstrated
that this radial range contains two stellar substructures having exponential
surface brightness profiles with different scalelengths, the inner one with
relatively low $h$, and the outer one with more extended (standard for giant
galaxies) scalelength. Our analysis reveals that the inner component has the
scaleheight of 0.2--0.7~kpc and the outer one about 1--3~kpc. Hence, we can
classify the inner and outer components as thin and thick stellar discs
respectively.

Having applied the full spectral fitting {\sc NBursts} technique with
a grid of high-resolution SSP models, we deduced mean ages and metallicities
of stars. The inner disc exhibits the intermediate age of about 5~Gyr and
the mean metallicity of $-0.2$~dex with a strong negative metallicity
gradient along the radius. The outer disc is quite metal-poor
([Fe/H]$=-0.4$~dex) and relatively young (2~Gyr). The outer disc harbours a
prominent starforming ring with the mean age of about 1~Gyr at the radius of
$R \approx 75\arcsec$ very well visible in UV data from the GALEX satellite
and 8$\mu$m NIR data from the Spitzer Space Telescope. Interestingly, in
this ring the mean stellar metallicity falls down to a very low value of
$-0.7$~dex. The galaxy nucleus and the bulge of NGC~7217 inside the nuclear
starforming ring are very old ($\sim 13-15$~Gyr).

How can we explain the complex structure of NGC~7217? In principle, the fact
that the outer stellar disc is younger than the inner one is in line with
the paradigm of the `inside-out' disc formation. But the sharp boundary
between the two discs, where the properties including stellar age and
metallicity as well as the radial and vertical scalelengths of the star
distribution change abruptly, hints to a temporal gap between the inner and
outer disc formation, or to some catastrophic event that provoked the
expansion of star formation into the outer ring area.

\begin{figure}
\includegraphics[height=\hsize,angle=270]{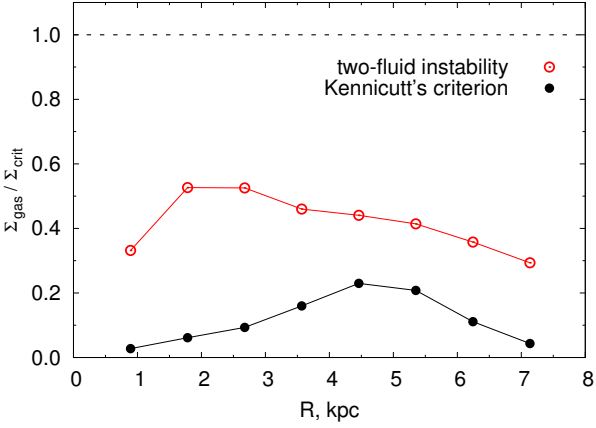}
\caption{Ratio between observed gas density and critical density according
to  \citet{Kennicutt89} 
and the two-fluid instability \citep{JS84,Efstathiou00} criteria.
The dashed line indicates the threshold for star formation. The H{\sc i}
surface density profile was taken from \citet{Noordermeer+05}.
\label{fig_cr}}
\end{figure}

Interestingly, the molecular gas concentrates in the inner disc
\citep{Combes+04} while the neutral hydrogen is observed in the narrow ring
in the outer disc \citep{VBA95,Buta+95,Noordermeer+05}. However, star
formation is almost absent in the inner disc (though it burns in the nuclear
starforming ring between the bulge and the inner disc) and is very prominent
in the outer disc \citep{Battinelli+00}.  \citet{Noordermeer+05}
estimated the gravitational stability of the gas in the outer ring and the
threshold density for star formation according to the \citet{Kennicutt89}
criterion,
\begin{equation}
\Sigma_\mathrm{cr} = \alpha \, \frac{\kappa c}{3.36 G} \, ,
\end{equation}
where $\kappa$ is the epicyclic frequency, that can be derived from 
the rotation and $c$ is the gas velocity dispersion, which can 
be assumed as a constant value of $6$~km~s$^{-1}$. For the dimensionless quantity 
$\alpha$, Kennicutt derived a value of 0.67 from the empirical study 
of star formation cutoffs in spiral galaxies. Dynamically, it means 
that one should take into account non-axisymmetric modes while 
considering the stability of a disc, because for axisymmetric modes 
$\alpha = 1$ \citep{Toomre64}.
\citet{Noordermeer+05} demonstrated that even in the middle of the H{\sc i} 
ring of NGC~7217, the gas density does not exceed 25~per~cent of the critical one. 
The estimate by \citet{Noordermeer+05} did not take into account the 
existence of a stellar disc. Having restored the velocity dispersion 
profile $\sigma_R$ for stars, we can evaluate the gaseous disc stability 
in the framework of the two-fluid approach 
\citep{JS84,Efstathiou00}
\begin{equation}
\Sigma_\mathrm{cr} = \frac{\kappa c}{3.36 G g(a,b)} \, ,
\end{equation}
where $a$ and $b$ are the ratios of stellar to gas velocity dispersion and
surface densities respectively, and $g(a,b)$ is a function derived
numerically by \citet{Efstathiou00}. This approach results in the
critical density estimate at least twice lower than the value obtained by
\citet{Noordermeer+05}. But even now the critical density remains too
high (see Fig.~\ref{fig_cr}). 
Therefore, the gas in the outer ring must be gravitationally stable,
and why the star formation proceeds there is still an open question. The very
existence of a system of starforming, gas-populated rings in an unbarred
galaxy is a puzzle. \citet{Combes+04} discussed a possibility of a past bar,
strong or weak, which had formed the rings and then had dissolved after the
gas inflow had supplied mass into the center. Now we see that this scenario
experiences strong difficulties: the stellar population in the NGC~7217
centre is very old, and it is clear that no significant star formation took
place there for the last 5~Gyr in order to provide mass concentration and
the subsequent bar destruction.

\begin{figure}
\includegraphics[width=\hsize]{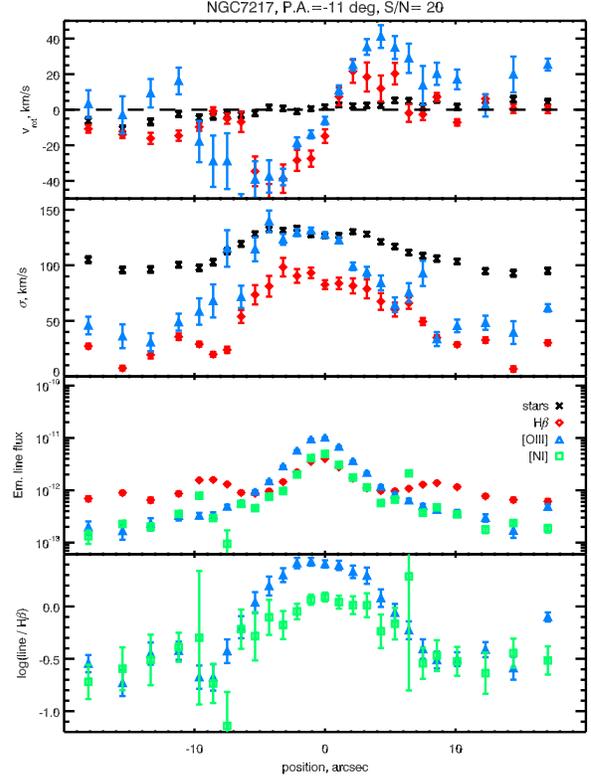}
\caption{Minor-axis kinematics and emission
line properties of NGC~7217 from the analysis of the 
H$\beta$ (red, diamonds), 
[O{\sc iii}] ($\lambda = 5007$~\AA, blue, triangles), and 
[N{\sc i}] ($\lambda = 5197/5200$~\AA, green, squares) 
lines in its spectra with the best-fitting
stellar population models subtracted. The panels show (top to bottom):
radial velocities, velocity dispersion, line fluxes, logarithms of the 
emission line ratios: [O{\sc iii}]/H$\beta$ (blue) and [N{\sc i}]/H$\beta$
(green).\label{minaxemfig}}
\end{figure}

\begin{figure}
\centering
\includegraphics[width=\hsize]{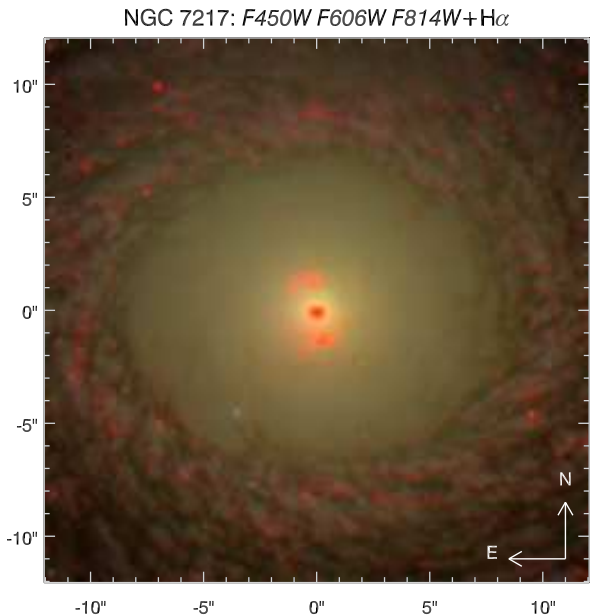}
\caption{Inner polar disc in NGC~7217 as seen with the HST. This false
colour composite is made of $F450W$, $F606W$, and $F814W$ WFPC2 images with
the H$\alpha$ fluxes from the ACS $F658N$ images added to the red channel
using the \citet{Lupton+04} algorithm.  The strongly inclined gaseous
structure seen here in the centre is a polar disc because the outer disc 
is seen almost face-on with its major axis nearly horizonthal in this plot.}
\label{poldischstfig}
\end{figure}

\subsection{Inner polar disc}

In Fig.~\ref{minaxemfig} we present the minor-axis kinematics and
emission line properties in the central region of NGC~7217 derived from the
analysis of the stellar population spectral fitting residuals. 
In the nuclear region
the [N{\sc i}] doublet is clearly detected in addition to the oxygen and
hydrogen lines. The [O{\sc iii}] kinematics is evident of either a rotating
disc or an outflow from the galaxy centre which can be induced by an active
nucleus. We measured the emission line ratios (see bottom panel in
Fig.~\ref{minaxemfig}) and used the {\sc itera} tool \citep{GA10} 
providing a large collection of emission line models for different
excitation mechanisms to perform the diagnostics. The extremely high
$\log($[N{\sc i}]/H$\beta) > 0$ ratio given moderate $\log($[O{\sc
iii}]/H$\beta) \approx 0.45$ is consistent with either a shock ionisation
for solar or subsolar metallicities of the ISM, or an AGN having a significantly
supersolar metallicity (e.g. $>+0.3$~dex). Given very moderate metallicity 
of the stellar population in the central region of NGC~7217, the latter possibility
can be ruled out at a high level of confidence. Therefore, we are probably
observing an inner polar disc in NGC~7217, which is supported by the direct
inspection of the HST imaging data in the narrow H$\alpha
+$[NII]$\lambda$6583-centered filter (Fig.~\ref{poldischstfig}).

According to \citet{Karachentseva73}, NGC~7217 is an isolated galaxy,
not a member of any group, and its environment is empty of satellites.
However, \citet{MK94} and \citet{SA00} suggested the presence of a counterrotating
stellar component in the central region of NGC~7217; moreover, we see also the
inner polar gaseous disc. Hence, accretion or minor merger(s) in the past
seem inevitable. If we assume that the galaxy centre is its oldest part then
we can suspect the vertical impact by a gas-rich satellite. Then the 0.3~kpc
nuclear polar disc in NGC~7217 is a remnant of the satellite gaseous
component which conserved its initial orbital momentum. The close analogue
of such an event is the Sagittarius dSph disruption by the Galaxy
accompanied by the re-distribution of the satellite material along the polar
large circle. Polar orbits are known to be stable, hence this gas cannot
accrete to the very centre explaining why the fuel for the star formation in
the centre of NGC~7217 has been absent until now evidenced by the old
stellar population there.

Such a vertical impact is also able to produce a system of
starforming/stellar rings by exciting a running compression wave in a large
scale galaxy disc \citep{APB97}. Moreover, the simulations of a
Cartwheel-like galaxy by \citet{Mapelli+08} for 1.5~Gyr after the impact
clearly show that sharp stellar rings formed initially expand over a large
area during 1~Gyr and form a structure resembling a low surface brightness
outer stellar disc. The two-tiered exponential stellar disc of NGC~7217
could have been built by such a minor merger. But we still need a
triaxiality of the potential in order to produce a counterrotating component
in the inner disc during the same event: the polar gas in the tumbling triaxial potential 
has to warp in the outer part in such a sense that it arrives to the rotational
plane of stars with the opposite spin \citep{vAKS82}. Being compressed, this
gas would be consumed by star formation leaving a counterrotating stellar
component. Besides, the triaxial potential also creates long-living
resonance rings in contrast to short-living collisional rings. From our data
we cannot distinguish between a triaxial dark-matter halo and a low-contrast
extended stellar spheroid (see discussion in \citealp{Buta+95}).

\subsection{Insights from simulations}

Simulations of vertical impacts producing ring structures reveal that the
vertical oscillations of disc stars excited by the intruder's passage result
in strong thickening of the stellar disc \citep{HW93} which we probably
observe in the outer disc of NGC~7217. The small thickness of the inner disc
can be perhaps explained by its high density increased after the impact by
the counterrotating component and also by the presence of the significant
molecular gas content \citep{Combes+04}. Interestingly, in the recognized
vertical collision product, the famous Cartwheel galaxy, the molecular gas
concentrates in the central region while the neutral atomic hydrogen is
confined to the outer ring \citep{Horellou+98} being very similar to
NGC~7217. 

One can suggest another explanation of the inner disc
structure. It could have been formed from the material of a small satellite
during its complete disruption \citep{EBAG06}. This new inner disc can be
older than the outer one because the infalling stars were initially older.
It can also be dynamically cold because its material had experienced the
orbital circularisation. However, this scenario requires a nearly in-plane
encounter. The unusually high thickness of the outer component and the
presence of the inner polar disc in  NGC~7217 make us to prefer encounters at higher
inclinations.

The minor merger scenario with a vertical impact requires this event to have
occurred long time ago. The galaxy needs time to convert the counterrotating
gas into stars and to establish the mean stellar age of 5~Gyr in the compact
high-density inner stellar disc. Then the current star formation in the
outer disc involving very low-metallicity gas does not relate to this event
and to its collision rings. It may be provoked by later gas infall in
a tidal tail developped during the minor merger. This mechanism can explain
the patchy and filamentary edges of the H{\sc i} ring in the map of NGC~7217
\citep{Noordermeer+05} and the presence of the shock-induced star formation
in the low-density gas ring which had to be gravitationally stable under the
quiescent conditions.

We explored intermediate resolution (0.2~kpc) TreeSPH simulations of
minor mergers provided by the GalMer
database\footnote{http://galmer.obspm.fr/} \citep{Chilingarian+10}.
Presently available simulations include interactions of a gas free giant
lenticular galaxy (``gS0'') with a 10 times less massive dwarf galaxy. The
morphology of a dwarf spans the entire Hubble sequence from non-rotating
ellipticals (``dE0'') to bulgeless discs (``dSd''). Here we consider only
gS0--dSd mergers, as this configuration has the largest quantity of gas
compared to other morphologies of a satellite galaxy. The inclination of the
orbital plane to the rotation planes of the two interacting galaxies is
33~deg for a giant S0 and 130~deg for dSd. The two sets of numerical
experiments include interactions on prograde and retrograde orbits 
with different initial orbital momenta and motion energies (24 in total). The
simulations were run for a total duration of 3~Gyr. 

The bar is always formed in the gS0 stellar disc. Merger remnant
morphologies vary quite a lot, but all of them can be classified into two
families related to the mutual orientation of the infalling satellite
internal angular momentum with respect to that of the orbital motion.

\begin{figure*}
\includegraphics[width=0.49\hsize]{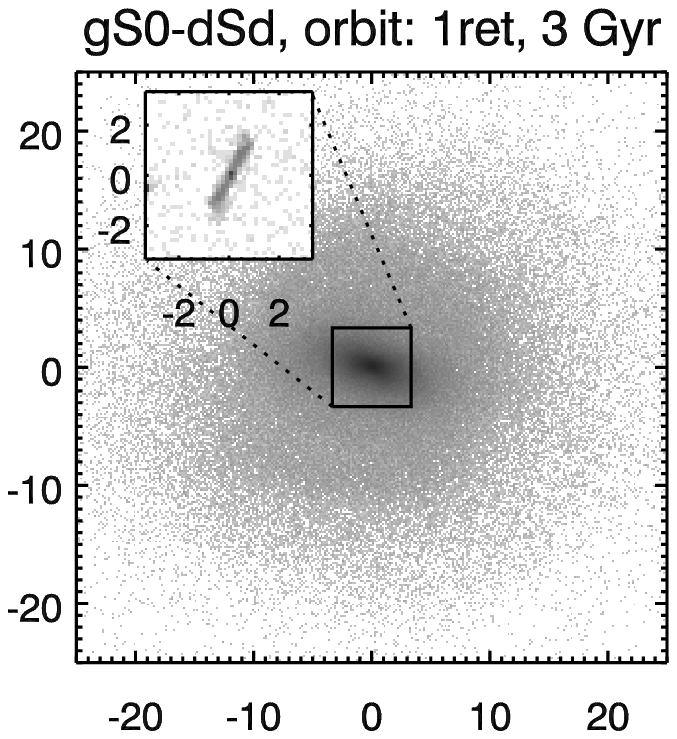}
\includegraphics[width=0.49\hsize]{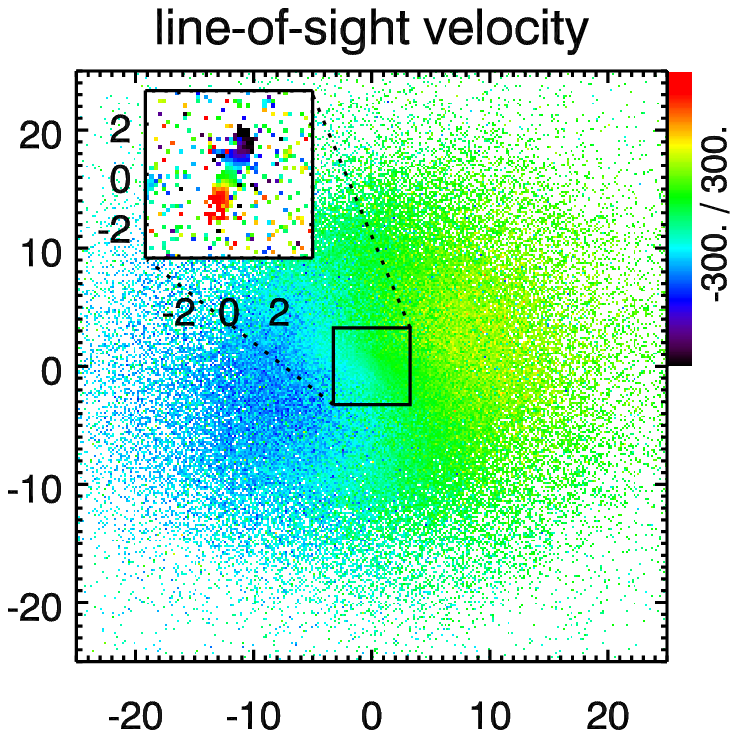}
\caption{\textit{Left panel:} surface density of stars and gas (inset) for a
minor wet merger of gS0 and dSd galaxies (mass ratio 10:1) on a retrograde
orbit. \textit{Right panel:} line-of-sight velocities of stars and gas
(inset) for the same merger. The signature of the inner polar ring seen
edge-on is clearly visible. The orientation of the galaxy disc is chosen to
match that of NGC~7217 ($i=30$~deg). The axes are in kpc.\label{figGalMerret}} 
\end{figure*}

All retrograde encounters result in the formation of a strongly
inclined inner starforming ring in the gS0 galaxy having a radius of
$\approx 1$~kpc. Its plane is orthogonal to the initial orbital plane of an
interaction, i.e. its inclination to the gS0 disc plane is about 60~deg.
This ring is located in the inner bright part of the S0 bulge and its
contribution to the total stellar light (and, consequently, to the stellar
kinematics derived from the spectra) in the central region of the galaxy is
negligible (see left panel of Fig.~\ref{figGalMerret}). However, it is
immediately revealed by the gas kinematics (see right panel of
Fig.~\ref{figGalMerret}), as it is the only structure with significant gas
content. Although the mass ratio of 1:10 in {\it total mass} is quite high,
none of the retrograde encounters heated up significantly the large scale S0
disc.

\begin{figure}
\includegraphics[width=\hsize]{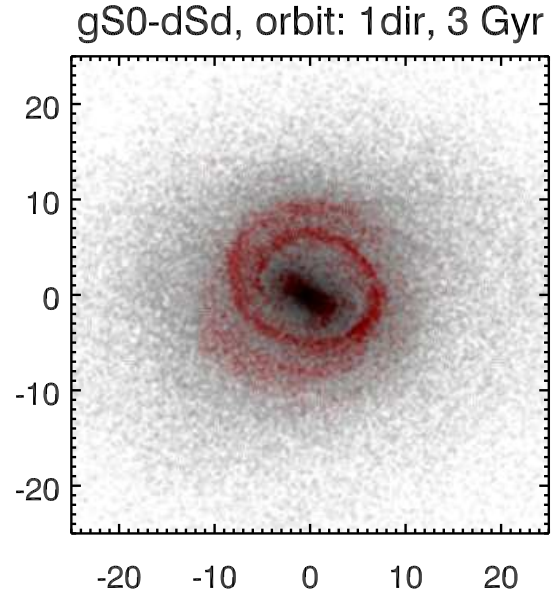}
\caption{Surface density of stars (grayscale) and gas (red) for a 
minor wet merger of gS0 and dSd galaxies (mass ratio 10:1) on a prograde
(direct) orbit. The orientation of the galaxy disc is chosen to match that
of NGC~7217 ($i=30$~deg). The axes are in kpc.\label{figGalMerdir}}
\end{figure}

\begin{figure}
\includegraphics[width=\hsize]{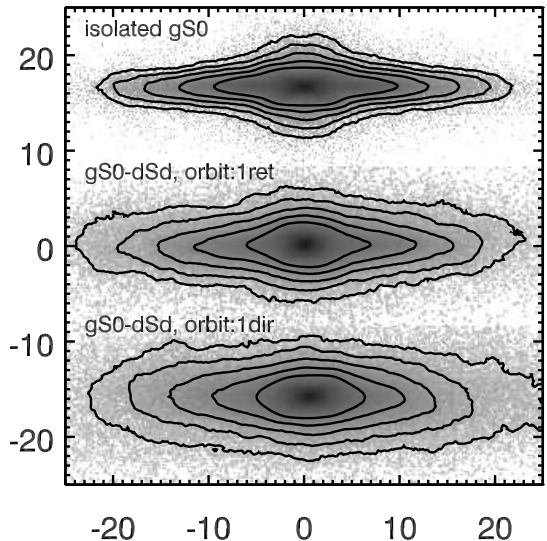}
\caption{Disc thickening by minor mergers. The edge-on view of the
GalMer model of a giant S0 galaxy in isolation is shown on top, 
the remnants of minor mergers with a gas-rich dSd galaxy (mass ratio 10:1,
orbit type 1, see text) on retrograde and prograde orbits are displayed in
the middle and in the bottom respectively. The contours are every
1~mag~arcsec$^{-2}$, the axes are in kpc.\label{figGalMerdisc}}
\end{figure}

The situation is totally different for prograde encounters. 
They heat and thicken the gS0 disc significantly, nearly destroying it for 
certain orbital types with high motion energies. The gas and newly formed 
stars from the infalling satellite form a large scale co-rotating disc in 
the main plane of the gS0 galaxy or between its initial plane orientation 
and the orbital plane if the large scale stellar disc is destroyed. If the 
large disc survives, the gas is concentrated at one of its resonances, 
probably the outer bar resonance, creating a starforming ring having 
$R \approx 8$~kpc quite thick ($\sim 1$~kpc) in the radial direction (see
Fig.~\ref{figGalMerdir}). As it resides in a region where the surface
density of the gS0 disc is rather low, it creates significant contribution
to the galaxy light. Since the initial metallicity of the interstellar
medium in dSd is low, the newly formed stars in this ring will also have low
metallicities being in agreement with young metal poor stars detected by our
stellar population analysis in the outer ring of NGC~7217. Worth mentioning
that due to the intense heating, the large scale gS0 disc flares in the
outer regions of a galaxy (see Fig~\ref{figGalMerdisc}). The outer
disc must have formed by this event not very long time ago, otherwise we
would have expected to see young metal-poor stars formed in it migrating
inwards and erasing an observed quite sharp break in the age profile. The
characteristic migration time of about 1~Gyr in a Milky Way-like galaxy
using the mechanism proposed recently by \citet{MF10} and \citet{Minchev+11}
allows us to constraint the formation epoch of the outer disc of NGC~7217.

Hence, we conclude that the scenario of two consequent minor mergers
at different orbits provides a plausible explanation to the observed
structure of NGC~7217.

\section{Summary}

We performed the analysis of internal kinematics and stellar
population properties in the nearby isolated early-type disc galaxy NGC~7217
having star-forming rings by the {\sc NBursts} full spectral fitting of
deep long-slit intermediate resolution spectroscopic data.

The age and metallicity profiles of NGC~7217 exhibit three components
that well correspond to those found in the structural analysis of the
galaxy by \citet{SA00}: an old metal-rich bulge dominating at radii
0.4--1.2~kpc, the inner disc with the mean age of 5~Gyr and slightly
subsolar metallicity at 1.6--4~kpc, and the outer metal-poor relatively
young disc ($t=$2--3~Gyr) beyond 4.8~kpc.

We analysed the parametrized major- and minor-axis kinematical
profiles of NGC~7217, recovered the velocity ellipsoid in the regions
dominated by the exponential discs and obtained the scaleheight radial
profile. We concluded that the inner disc is thin ($z_0=0.2\dots0.7$~kpc)
while the outer disc flares reaching the half-thickness of 2.5~kpc in the
region of the outer starforming ring. Using the two-fluid approach, we
evaluated the gaseous disc stability in NGC~7217 and showed that even the highest-density
neutral-hydrogen ring at $R=6$~kpc should be stable.

From the analysis of emission line ratios in the residuals of the
spectral fitting of NGC~7217, we ruled out a strong nuclear activity as
the mechanism of circumnuclear gas excitation and
concluded that the peculiar ionized-gas kinematics in the central region of
NGC~7217 was due to the presence of the inner polar disc.

By comparing our data with the numerical simulations of wet minor
mergers in the GalMer database, we explained the observed structure and
kinematics of NGC~7217 by the two consequent minor mergers with satellites
having different initial orbits. The encounter on a retrograde orbit resulted
in the formation of the inner polar disc, while the outer star-forming ring
and the outer disc thickening were consistent with a minor merger on a
prograde orbit.

\section*{Acknowledgments}

We thank Alexei Moiseev for the useful discussions. We are indebted to Alister
Graham for his consultation. We are also
grateful to our anonymous referee whose suggestions helped us to improve the
clarity and presentation of the results. This research is partly based on
observations made with the NASA/ESA Hubble Space Telescope, obtained from
the data archive at the Space Telescope Science Institute, which is operated
by the Association of Universities for Research in Astronomy, Inc., under
NASA contract NAS 5-26555. The work on the study of young stellar rings in
disc galaxies is supported by the grants of the Russian Foundation for Basic
Researches number 10-02-00062a and number 09-02-00968a.

\bibliographystyle{mn2e}
\bibliography{ngc7217}

\label{lastpage}

\end{document}